\documentclass[letterpaper, 10 pt, conference]{ieeeconf}
\IEEEoverridecommandlockouts
\usepackage{amsmath,amssymb,amsfonts}
\usepackage{mathtools}
\usepackage{graphicx}
\usepackage{textcomp}
\usepackage{xcolor}
\usepackage{booktabs}
\usepackage{multirow}
\usepackage{hyperref}

\newcommand{\etal}{et al.}

\newcommand{\eg}{e.g.}
\newcommand{\wrt}{w.r.t.}

\begin{document}

\title{\LARGE \bf IF-CPS: Influence Functions for Cyber-Physical Systems---\\A Unified Framework for Diagnosis, Curation, and Safety Attribution}

\author{Jiachen Li$^{1}$, Shihao Li$^{1}$, Soovadeep Bakshi$^{1}$, Jiamin Xu$^{1}$, and Dongmei Chen$^{1}$%
\thanks{$^{1}$All authors are with the Department of Mechanical Engineering, The University of Texas at Austin, Austin, TX 78712, USA.
{\tt\small \{jiachenli, shihaoli01301, soovadeepbakshi, jiaminxu\}@utexas.edu, dmchen@me.utexas.edu}}%
}

\maketitle

\begin{abstract}
Neural network controllers trained via behavior cloning are increasingly deployed in cyber-physical systems (CPS), yet practitioners lack tools to trace controller failures back to training data.
Existing data attribution methods assume i.i.d.\ data and standard loss targets, ignoring CPS-specific properties: closed-loop dynamics, safety constraints, and temporal trajectory structure.
We propose IF-CPS, a modular influence function framework with three CPS-adapted variants: safety influence (attributing constraint violations), trajectory influence (temporal discounting over trajectories), and propagated influence (tracing effects through plant dynamics).
We evaluate IF-CPS on six benchmarks across diagnosis, curation, and safety attribution tasks.
IF-CPS improves over standard influence functions in the majority of settings, achieving AUROC $1.00$ in Pendulum (5--10\% poisoning), $0.92$ vs.\ $0.50$ in HVAC (10\%), and the strongest constraint-boundary correlation (Spearman $\rho = 0.55$ in Pendulum).
\end{abstract}

%% ============================================================
\section{Introduction}
\label{sec:intro}

Neural network controllers trained on offline demonstrations are increasingly used in cyber-physical systems (CPS) such as autonomous vehicles~\cite{bojarski2016end}, robotic manipulators~\cite{chi2024universal}, and industrial process control.
Behavior cloning is a common approach due to its simplicity and data efficiency~\cite{pomerleau1991efficient}.
However, when a cloned controller exhibits unsafe behavior in deployment, practitioners face a fundamental question: which training demonstrations caused this failure?
Answering this question is essential for diagnosing failures, curating training datasets, and building safety cases for controller certification.

Data attribution methods from machine learning offer a principled answer.
Influence functions~\cite{koh2017understanding} approximate the effect of removing a training point via a Hessian-inverse-gradient product.
LiSSA~\cite{agarwal2017second} makes this feasible for deep networks; TracIn~\cite{pruthi2020tracin} uses gradient dot products across checkpoints; TRAK~\cite{park2023trak} uses random projections for scalability; Data Shapley~\cite{ghorbani2019data} provides game-theoretic valuation; and Choe~\etal~\cite{choe2025logra} and Hammoudeh and Lowd~\cite{hammoudeh2024survey} address large-scale attribution and provide comprehensive surveys.
These methods have proven effective for classification tasks, but their application to CPS control remains largely unexplored.

The challenge is that CPS controllers differ from standard supervised learning in three fundamental ways.
First, closed-loop dynamics mean the controller's output feeds back through the physical plant, so a training point's influence propagates through system dynamics rather than affecting a single prediction.
Second, safety constraints are often the primary concern in CPS---constraint satisfaction (\eg, staying within position or temperature bounds) matters more than loss minimization alone.
Third, temporal trajectory structure means training demonstrations are trajectories, not independent samples, so errors at early timesteps compound through subsequent states.

Recent work has begun extending data attribution beyond the standard supervised learning setting.
In the reinforcement learning domain, Abolfazli~\etal~\cite{dvorl2022} propose data valuation for offline RL, while Hu~\etal~\cite{snapshot2025} adapt TracIn-based attribution to the online RL setting.
However, neither line of work addresses CPS-specific challenges such as safety constraints or closed-loop dynamics propagation.
In the controls domain, Li~\etal~\cite{li2025influence} derive influence functions for linear system identification and LQR control through DARE sensitivity analysis, but their formulation is restricted to linear systems and does not accommodate safety constraints or neural network policies.
CUPID~\cite{nishimura2025cupid} applies influence functions to robot imitation learning for data curation, yet does not explicitly target safety-critical CPS requirements.
While existing neural network verification tools can certify controller safety using set-based reachability methods, they do not trace safety violations back to individual training demonstrations.
IF-CPS complements such verification approaches by providing data-level explanations for why a learned controller satisfies or violates safety specifications.
We adopt standard IF~\cite{koh2017understanding} as the primary baseline, as it is the direct parent method from which IF-CPS derives; this first CPS-focused study is designed to isolate the value of domain-specific adaptations before benchmarking against broader data valuation families.

\begin{figure}[t]
\centering
\includegraphics[width=\columnwidth]{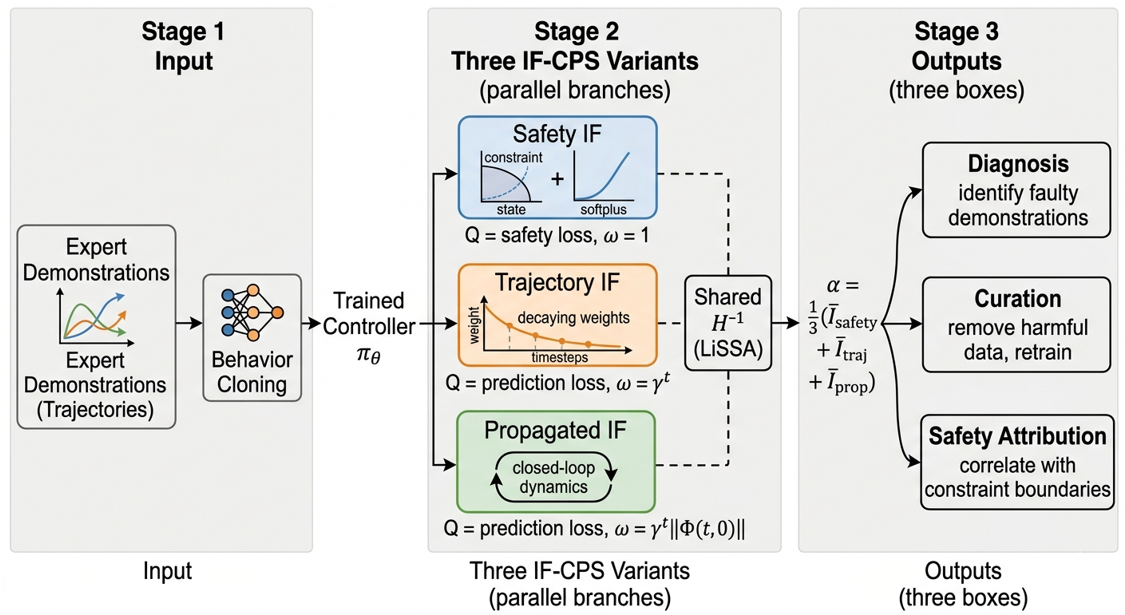}
\caption{Overview of the IF-CPS framework. }
\label{fig:overview}
\end{figure}

We propose IF-CPS, a modular influence function framework that adapts data attribution to the CPS domain through three variants---safety influence, trajectory influence, and propagated influence---supporting diagnosis, curation, and safety attribution of learned controllers (Fig.~\ref{fig:overview}).
Our contributions are: (1) three CPS-adapted influence function variants that account for closed-loop dynamics, safety constraints, and temporal trajectory structure; (2) a modular framework with a shared inverse-Hessian computation, where variants can be applied individually or combined as a parameter-free ensemble; and (3) evaluation on six CPS benchmarks (including Quadrotor figure-8 tracking, multi-zone HVAC, and CSTR chemical process control) showing improvements over standard influence functions in the majority of settings, with ablation identifying which variant contributes most in each domain.

%% ============================================================
\section{Problem Formulation}
\label{sec:problem}

\subsection{CPS Control via Behavior Cloning}

We consider a discrete-time CPS with state $x \in \mathcal{X} \subseteq \mathbb{R}^n$, action $u \in \mathcal{U} \subseteq \mathbb{R}^m$, and known plant dynamics $x_{t+1} = f(x_t, u_t)$.
A neural network controller $\pi_\theta: \mathcal{X} \rightarrow \mathcal{U}$ is trained via behavior cloning on a dataset of expert demonstrations $\mathcal{D} = \{\tau_1, \ldots, \tau_N\}$, where each demonstration $\tau_i = \{(x^i_0, u^i_0), (x^i_1, u^i_1), \ldots, (x^i_{T_i}, u^i_{T_i})\}$ is a state-action trajectory.
The training objective minimizes:
\begin{equation}
    L(\theta) = \frac{1}{|\mathcal{D}|} \sum_{\tau_i \in \mathcal{D}} \sum_{t=0}^{T_i} \ell(\pi_\theta(x^i_t), u^i_t)
\end{equation}
where $\ell$ is a per-step loss (\eg, cross-entropy or MSE).

Flattening the trajectory structure, define the full training set of state-action pairs $\mathcal{D}_{flat} = \{(x_j, u_j)\}_{j=1}^{M}$ where $M = \sum_{i=1}^{N} (T_i + 1)$.
The empirical risk is:
\begin{equation}
    L(\theta) = \frac{1}{M} \sum_{j=1}^{M} \ell(\pi_\theta(x_j), u_j)
    \label{eq:empirical_risk}
\end{equation}
For continuous actions (MSE loss): $\ell(\pi_\theta(x), u) = \frac{1}{m} \| \pi_\theta(x) - u \|_2^2$.
For discrete actions (cross-entropy on logits $z_\theta(x) \in \mathbb{R}^{|\mathcal{U}|}$): $\ell(\pi_\theta(x), u) = -z_\theta(x)_u + \log \sum_{a \in \mathcal{U}} \exp(z_\theta(x)_a)$, where $z_\theta(x)_u$ denotes the logit corresponding to the expert action $u$.

\subsection{Safety Constraints}

The CPS has $K$ safety constraints $g_k(x) \leq 0$ for $k = 1, \ldots, K$.
A state $x$ is safe if all constraints are satisfied.
A trajectory $\tau$ has a safety violation at time $t$ if $\exists k: g_k(x_t) > 0$.

Define the safe set $\mathcal{S} = \{x \in \mathcal{X} : g_k(x) \leq 0,\; \forall k = 1, \ldots, K\}$ and the constraint boundary $\partial \mathcal{S}_k = \{x : g_k(x) = 0\}$.
The signed distance to the nearest constraint boundary at state $x \in \mathcal{S}$ is:
\begin{equation}
    d(x) = \min_{k=1,\ldots,K} \frac{-g_k(x)}{\|\nabla_x g_k(x)\|_2}
    \label{eq:signed_distance}
\end{equation}
which is well-defined when $\nabla_x g_k(x) \neq 0$ on $\partial \mathcal{S}_k$ (constraint qualification).
The maximum constraint violation along a trajectory $\tau$ is $V(\tau) = \max_{t=0,\ldots,T} \max_{k=1,\ldots,K} g_k(x_t)$.
A trajectory is safe iff $V(\tau) \leq 0$.

\subsection{Data Attribution Problem}

Given a trained controller $\pi_{\hat{\theta}}$ and a test scenario (trajectory $\tau_{test}$) where the controller violates safety constraints, the data attribution problem is to assign a score $\alpha_i \in \mathbb{R}$ to each training demonstration $\tau_i$ quantifying its responsibility for the violation.

%% ============================================================
\section{IF-CPS: Methodology}
\label{sec:method}

\subsection{Standard Influence Function }

We briefly review the derivation to establish notation.
Consider upweighting a single training point $z_i = (x_i, u_i)$ by $\epsilon$, yielding the perturbed parameters:
\begin{equation}
    \hat{\theta}_\epsilon = \arg\min_\theta \; L(\theta) + \epsilon \, \ell(\pi_\theta(x_i), u_i)
\end{equation}
By the implicit function theorem applied to the first-order optimality condition $\nabla_\theta L(\hat{\theta}_\epsilon) + \epsilon \nabla_\theta \ell(z_i, \hat{\theta}_\epsilon) = 0$, differentiating \wrt\ $\epsilon$ at $\epsilon = 0$:
\begin{equation}
    \frac{d \hat{\theta}_\epsilon}{d \epsilon}\bigg|_{\epsilon=0} = -H_{\hat{\theta}}^{-1} \nabla_\theta \ell(z_i, \hat{\theta})
    \label{eq:param_influence}
\end{equation}
where $H_{\hat{\theta}} = \nabla^2_\theta L(\hat{\theta})$ is the Hessian of the training loss.
The influence of $z_i$ on any differentiable test quantity $\mathcal{Q}(\theta)$ is then:
\begin{equation}
    \mathcal{I}(z_i, \mathcal{Q}) = -\nabla_\theta \mathcal{Q}(\hat{\theta})^\top H_{\hat{\theta}}^{-1} \nabla_\theta \ell(z_i, \hat{\theta})
    \label{eq:general_if}
\end{equation}
The standard IF sets $\mathcal{Q}(\theta) = \ell(\pi_\theta(x_{test}), u_{test})$, the prediction loss at a test point.
IF-CPS defines three alternative test objectives $\mathcal{Q}$ suited to CPS applications.

\subsection{Safety Influence Function}

Standard influence functions attribute changes in prediction loss.
For CPS, we instead attribute safety violations.
Define the smoothed safety loss for a fixed test trajectory $\tau_{test}$ as:
\begin{equation}
    L_{safety}(\tau_{test}, \theta) = \sum_{t=0}^{T} \sum_{k=1}^{K} \phi_\beta(g_k(x^{test}_t, \theta))
    \label{eq:safety_loss}
\end{equation}
where $\phi_\beta(s) = \frac{1}{\beta}\log(1 + e^{\beta s})$ is the softplus function and $g_k(x^{test}_t, \theta) \coloneqq g_k\bigl(f(x^{test}_{t-1},\, \pi_\theta(x^{test}_{t-1}))\bigr)$ for propagated states, or simply $g_k(x^{test}_t)$ for the fixed-trajectory setting.
Setting $\mathcal{Q} = L_{safety}$ in~\eqref{eq:general_if}, the safety influence of training demonstration $\tau_i$ is:
\begin{equation}
    I_{safety}(\tau_i, \tau_{test}) = -\nabla_\theta L_{safety}(\tau_{test}, \hat{\theta})^\top H^{-1}_{\hat{\theta}} \nabla_\theta L(\tau_i, \hat{\theta})
    \label{eq:safety_if}
\end{equation}
This follows directly from~\eqref{eq:general_if}.
Since $\phi_\beta$ is $C^\infty$, $g_k$ is $C^1$, and $\pi_\theta$ is piecewise differentiable (ReLU), the chain rule gives:
\begin{equation}
    \nabla_\theta L_{safety} = \sum_{t,k} \underbrace{\sigma(\beta\, g_k(x^{test}_t))}_{\text{soft indicator}} \cdot \nabla_x g_k^\top \cdot \frac{\partial f}{\partial u}\bigg|_{x^{test}_t} \cdot \nabla_\theta \pi_\theta(x^{test}_t)
    \label{eq:safety_grad_chain}
\end{equation}
where $\sigma(s) = 1/(1+e^{-s})$ is the sigmoid function.
The sigmoid weighting $\sigma(\beta\, g_k(x))$ acts as a soft indicator for constraint violation: it approaches $1$ for violated states ($g_k > 0$), decays exponentially for safe states far from the boundary, and equals ${\approx}1/2$ near the boundary.
The effective sensitivity region has width $\Delta = 2/\beta$ centered on $\partial \mathcal{S}_k$, so the safety gradient is dominated by timesteps where the test trajectory is near a constraint boundary.

The smoothed safety loss is related to the log-barrier function used in constrained optimization.
As $\beta \to \infty$, $\phi_\beta(g_k) \to [g_k]_+$, recovering the exact constraint violation penalty.
The choice of $\beta$ controls a bias-variance trade-off: large $\beta$ gives precise but noisy attribution, while small $\beta$ gives smooth but less discriminative attribution.

We approximate $H^{-1}v$ using LiSSA~\cite{agarwal2017second}.

\subsection{Trajectory Influence}

Standard influence functions treat training data as i.i.d.\ samples.
In CPS, demonstrations are trajectories with temporal structure: errors at early timesteps compound through dynamics.
For a training demonstration $\tau_i$ and a single test state-action pair $(x^{test}_t, u^{test}_t)$, the per-timestep influence is:
\begin{equation}
    I_t(\tau_i) \coloneqq -\nabla_\theta \ell(\pi_{\hat{\theta}}(x^{test}_t), u^{test}_t)^\top H^{-1}_{\hat{\theta}} \nabla_\theta L(\tau_i, \hat{\theta})
\end{equation}
The trajectory influence aggregates these with geometric temporal discounting:
\begin{equation}
    I_{traj}(\tau_i, \tau_{test}) = \sum_{t=0}^{T} \gamma^t \cdot I_t(\tau_i)
    \label{eq:traj_if}
\end{equation}
where $\gamma \in (0, 1)$ is the discount factor.
Defining the discount-weighted test gradient $\tilde{g} = \sum_{t=0}^{T} \gamma^t \nabla_\theta \ell(\pi_{\hat{\theta}}(x^{test}_t), u^{test}_t)$, this reduces to:
\begin{equation}
    I_{traj}(\tau_i, \tau_{test}) = -\tilde{g}^\top H^{-1}_{\hat{\theta}} \nabla_\theta L(\tau_i, \hat{\theta})
\end{equation}
which is a standard IF with an aggregated test gradient, requiring only one LiSSA solve instead of $T+1$.
The normalized discount weights form a truncated geometric distribution with effective horizon:
\begin{equation}
    T_{eff} = \frac{\gamma}{1-\gamma} - \frac{(T+1)\gamma^{T+1}}{1 - \gamma^{T+1}}
    \label{eq:eff_horizon}
\end{equation}
For $\gamma = 0.99$ and $T = 200$: $T_{eff} \approx 85.7$ steps, placing 42\% of weight on the first half of the trajectory.

\subsection{Propagated Influence}

The key insight is that in CPS, a training point's effect propagates through the closed-loop system dynamics.
The closed-loop Jacobian at state-action pair $(x_t, u_t)$ along the test trajectory is:
\begin{equation}
    J_{cl}(t) = \underbrace{\frac{\partial f}{\partial x}\bigg|_{x_t, u_t}}_{A(t)} + \underbrace{\frac{\partial f}{\partial u}\bigg|_{x_t, u_t}}_{B(t)} \cdot \underbrace{\frac{\partial \pi_\theta}{\partial x}\bigg|_{x_t}}_{K_\theta(t)}
    \label{eq:cl_jacobian}
\end{equation}
The cumulative state transition from time $0$ to $t$ is $\Phi(t, 0) = \prod_{k=0}^{t-1} J_{cl}(k)$, with $\Phi(0, 0) = I_n$.
The spectral norm $\|\Phi(t, 0)\|_2$ measures worst-case perturbation amplification over $t$ timesteps.
The propagated influence weights per-timestep attributions by this dynamical sensitivity:
\begin{equation}
    I_{prop}(\tau_i, \tau_{test}) = \sum_{t=0}^{T} \gamma^t \, \|\Phi(t, 0)\|_2 \cdot I_t(\tau_i)
    \label{eq:prop_if}
\end{equation}

For linear plants ($f(x,u) = Ax + Bu$), $J_{cl} = A + B \frac{\partial \pi}{\partial x}$ is constant, and the propagation reduces to powers of a fixed matrix.
For nonlinear plants, we compute $J_{cl}(t)$ at each timestep via first-order Taylor expansion around the current operating point.
Under linear dynamics with constant controller Jacobian $K_\theta$, the closed-loop Jacobian is time-invariant: $J_{cl} = A_{cl} \coloneqq A + BK_\theta$.
If the closed-loop system is Schur stable ($\rho(A_{cl}) < 1$), there exist $C \geq 1$, $c \in [\rho(A_{cl}), 1)$ with $\|A_{cl}^t\|_2 \leq C c^t$, yielding the bound:
\begin{equation}
    |I_{prop}(\tau_i, \tau_{test})| \leq \frac{C}{1 - \gamma c} \cdot \max_{t} |I_t(\tau_i)|
    \label{eq:prop_bound}
\end{equation}
When $\rho(A_{cl}) \to 1^-$ (approaching the stability boundary), the total propagated IF weight diverges, automatically assigning larger attribution scores when the closed-loop system is near instability---a physically meaningful property, since a controller operating near its stability margin is most sensitive to training data quality.

This parallels the approach of Li~\etal~\cite{li2025influence}, whose IF2 computes influence on LQR cost through DARE sensitivity with contributions scaled by $\|(A_{cl}^*)^t\|_2^2$.
Both methods weight perturbation influence by powers of the closed-loop transition matrix---Li~\etal\ implicitly through the DARE algebraic structure (exact for LQR); IF-CPS explicitly as a scalar weight (applicable to nonlinear neural network controllers).

\textbf{Assumptions.} The propagated variant requires: (A1)~known or identified plant dynamics $f$ with available Jacobians $\partial f/\partial x$, $\partial f/\partial u$; (A2)~a differentiable controller (piecewise for ReLU networks); (A3)~finite propagation horizon to bound linearization error in nonlinear systems. The Schur stability bound~\eqref{eq:prop_bound} additionally requires a stable closed-loop system.
When dynamics are unavailable or poorly identified, practitioners should use the safety and trajectory variants only; the propagated variant is optional and adds value primarily when reliable dynamics models exist.

\subsection{Modular Framework}

All three IF-CPS variants share a common computational template, differing only in the per-timestep test objective $\mathcal{Q}_t$ and the timestep weight $\omega_t$:
\begin{equation}
    I_{CPS}(\tau_i, \tau_{test}) = -\biggl(\sum_{t=0}^{T} \omega_t \cdot \nabla_\theta \mathcal{Q}_t(\hat{\theta})\biggr)^\top H^{-1}_{\hat{\theta}} \nabla_\theta L(\tau_i, \hat{\theta})
    \label{eq:unified}
\end{equation}
where $\mathcal{Q}_t$ is the per-timestep test objective and $\omega_t$ is the timestep weight:

\begin{center}
\small
\begin{tabular}{lcc}
\toprule
Variant & $\mathcal{Q}_t(\theta)$ & $\omega_t$ \\
\midrule
Standard IF & $\ell(\pi_\theta(x^{test}_t), u^{test}_t)$ & $1/(T{+}1)$ \\
Safety IF   & $\sum_k \phi_\beta(g_k(x^{test}_t))$ & $1$ \\
Trajectory IF & $\ell(\pi_\theta(x^{test}_t), u^{test}_t)$ & $\gamma^t$ \\
Propagated IF & $\ell(\pi_\theta(x^{test}_t), u^{test}_t)$ & $\gamma^t \|\Phi(t,0)\|_2$ \\
\bottomrule
\end{tabular}
\end{center}

The combined IF-CPS score is computed as a parameter-free equal-weight ensemble:
\begin{equation}
    \alpha_i^{IF\text{-}CPS} = \frac{1}{3}\left[\bar{I}_{safety}(\tau_i) + \bar{I}_{traj}(\tau_i) + \bar{I}_{prop}(\tau_i)\right]
    \label{eq:combined}
\end{equation}
where $\bar{I}$ denotes min-max normalization to $[0, 1]$.
Equal weighting is a deliberate parameter-free default; no optimality claim is made for this aggregation rule, and practitioners may use individual variants when domain knowledge suggests one signal dominates (see ablation in Section~\ref{sec:experiments}).

Each IF-CPS variant requires exactly one inverse-Hessian-vector product $H^{-1}_{\hat{\theta}} \nabla_\theta L(\tau_i, \hat{\theta})$ per training demonstration, which can be shared across all variants.
Computing all three variants simultaneously costs approximately the same as computing one, plus $O(3 \cdot T \cdot p)$ for the test-side gradients and $O(T \cdot n^2 \cdot p)$ for the Jacobian products in the propagated variant.

\subsection{Approximation Quality Analysis}

The LiSSA algorithm approximates $H^{-1} v$ via the Neumann series:
\begin{equation}
    H^{-1} v \approx \sum_{r=0}^{R-1} (I - \alpha H)^r (\alpha v)
    \label{eq:lissa}
\end{equation}

With damped Hessian $\tilde{H} = H + \lambda I$, scale $\alpha = 1 / \tilde{\lambda}_p$, and $R$ recursion steps, the approximation error is bounded by:
\begin{equation}
    |I_{exact}(\tau_i) - I_{approx}(\tau_i)| \leq \|q\|_2 \cdot \frac{\|\nabla_\theta L(\tau_i)\|_2}{\lambda} \cdot \biggl(1 - \frac{\lambda}{\tilde{\lambda}_p}\biggr)^R
    \label{eq:if_error}
\end{equation}
where $q = \sum_t \omega_t \nabla_\theta \mathcal{Q}_t$ is the test-side gradient.
The relative ranking of training points (which AUROC measures) is more robust to uniform approximation error than absolute scores.

\subsection{Computational Complexity}

Each influence function computation requires one gradient computation per training trajectory ($O(N)$) and one LiSSA inverse-Hessian-vector product ($O(R \cdot B)$ for $R$ recursion steps and batch size $B$).
The propagated influence adds $O(T \cdot n^2)$ per trajectory for Jacobian computation ($n$ = state dimension, $T$ = propagation horizon).
Total complexity per test trajectory: $O(N \cdot (R \cdot B + T \cdot n^2))$.
For typical CPS datasets ($N \leq 200$, $n \leq 13$, $T \leq 50$), this is dominated by the LiSSA computation and remains tractable.

%% ============================================================
%% ============================================================
\section{Experiments}
\label{sec:experiments}

\subsection{Setup}

We evaluate on six CPS benchmarks (Table~\ref{tab:env_summary}): three standard Gymnasium~\cite{towers2024gymnasium} tasks (CartPole, Pendulum, LunarLander) and three domain-specific environments (Quadrotor figure-8 tracking with 6D state and 3D acceleration action; multi-zone HVAC thermal regulation with 4D state and comfort constraints $20 \leq T_i \leq 26$\,°C; CSTR chemical reactor with 3D state and safety constraints on temperature and concentration).

\begin{table}[htbp]
\caption{Evaluation environments summary.}
\label{tab:env_summary}
\centering
\small
\begin{tabular}{lcccl}
\toprule
Environment & $n$ & $m$ & Action & Safety Constraints \\
\midrule
CartPole    & 4 & 1 (disc.) & Discrete & $|\theta|, |x|$ bounds \\
Pendulum    & 3 & 1 (cont.) & Cont.    & $|\dot{\theta}|$ bound \\
LunarLander & 8 & 1 (disc.) & Discrete & $y > 0$, $|\theta|$ \\
Quadrotor   & 6 & 3 (cont.) & Cont.    & $z$, $|x|$, $|y|$ bounds \\
HVAC        & 4 & 1 (cont.) & Cont.    & $T_i \in [20, 26]$ \\
CSTR        & 3 & 1 (cont.) & Cont.    & $T_r, C_A$ bounds \\
\bottomrule
\end{tabular}
\end{table}

For each environment, we train a 2-layer MLP controller (64 hidden units, ReLU) via behavior cloning on 100 expert demonstrations, using Adam ($\text{learning rate} = 10^{-3}$), batch size 128, and early stopping (patience 10).
We compare Random (uniform scores), Standard IF~\cite{koh2017understanding}, and IF-CPS (Eq.~\ref{eq:combined}).
The empirical claim is improvement over standard IF baselines in CPS settings; comparison against broader data valuation methods (e.g., TRAK, Data Shapley) is left to future work.
We evaluate four protocols: (A) Diagnosis---inject faulty demonstrations at rates 5--20\% and measure AUROC; (B) Curation---retrain on top-ranked subsets; (C) Safety attribution---correlate attribution with constraint boundary proximity (Spearman $\rho$); (D) Ablation---compare IF-CPS variants.
All experiments use 3 seeds; LiSSA: 5 recursions, damping 0.01; propagation horizon 20 steps; $\gamma = 0.99$.

\subsection{Trajectory Visualization}

\begin{figure*}[htbp]
\centering
\includegraphics[width=\textwidth]{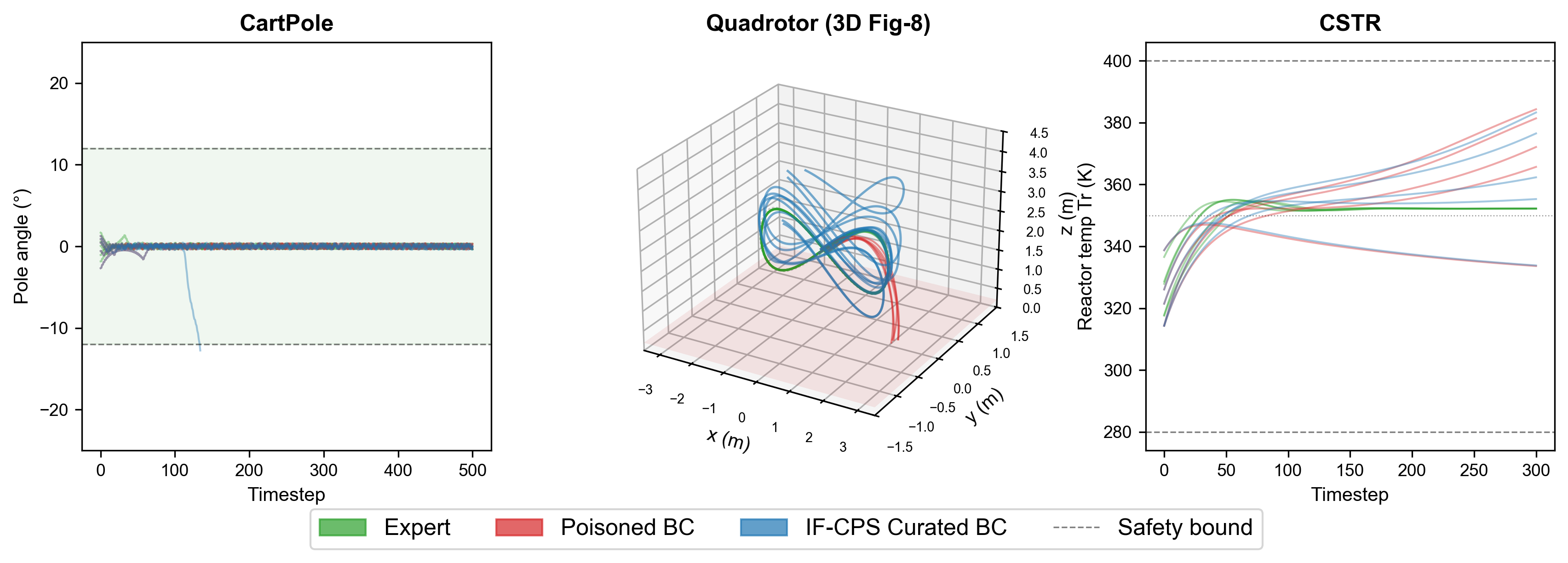}
\caption{IF-CPS data curation on CartPole, Quadrotor, and CSTR: expert (green), 20\% poisoned BC (red), IF-CPS-curated BC (blue). Dashed lines show safety bounds.}
\label{fig:trajectories}
\end{figure*}

Fig.~\ref{fig:trajectories} compares trajectories under the expert (green), 20\% poisoned BC (red), and IF-CPS-curated BC (blue, bottom 30\% removed).
In CartPole, curation recovers near-expert pole stabilization.
In Quadrotor, the poisoned BC diverges rapidly while curation recovers figure-8 tracking.
In CSTR, the poisoned BC overshoots toward the 400\,K safety bound; curation recovers smooth transient behavior.

\subsection{Diagnosis}

\begin{figure}[htbp]
\centering
\includegraphics[width=\columnwidth]{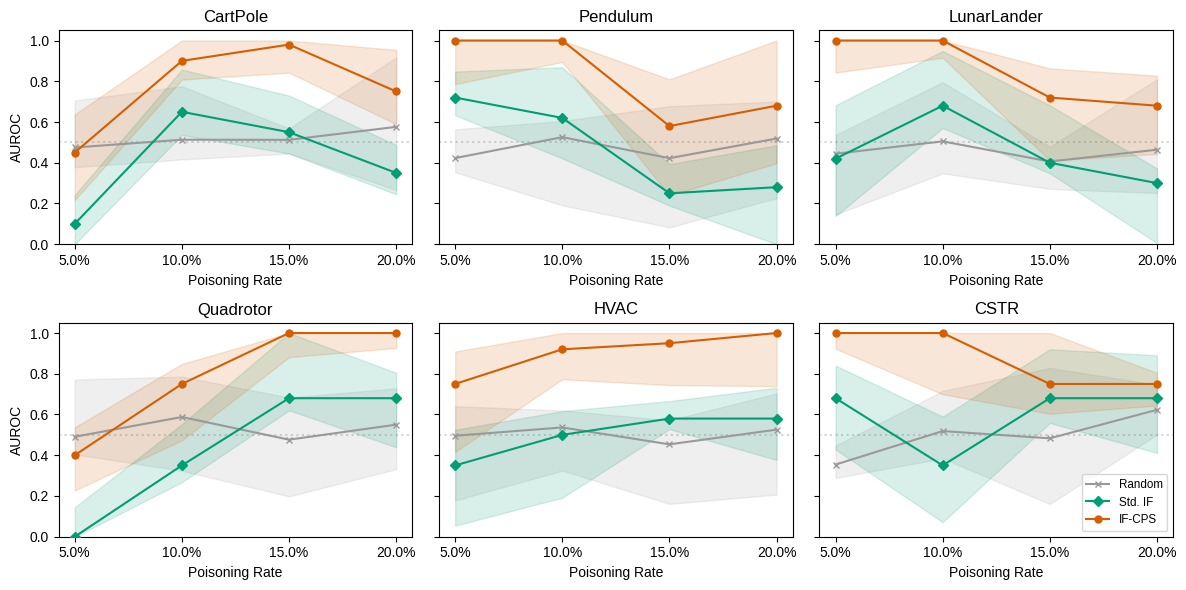}
\caption{Diagnosis AUROC vs.\ poisoning rate. Shaded regions: $\pm$1 std (3 seeds).}
\label{fig:diagnosis}
\end{figure}

Table~\ref{tab:diagnosis} and Fig.~\ref{fig:diagnosis} report AUROC for identifying faulty demonstrations.
As a descriptive summary across the 24 environment-rate combinations, IF-CPS outperforms Standard IF in 18 cases, ties in 2, and loses in 4 (win rate 75\%; note that with $n=3$ seeds these are point-estimate comparisons, not formal hypothesis tests).
The largest gains appear in Pendulum (AUROC $1.00$ vs.\ $0.72$ at 5\%), CartPole ($0.90$ vs.\ $0.63$ at 10\%), HVAC ($0.92$ vs.\ $0.50$ at 10\%), and CSTR ($0.73$ vs.\ $0.33$ at 10\%).
LunarLander at 10\% is a tie ($0.67$ vs.\ $0.68$), and several Quadrotor settings show high variance ($\pm 0.47$) due to the small seed count ($n=3$), indicating that these point estimates should be interpreted cautiously.
Performance tends to be strongest at 5--10\% poisoning rates and less consistent at 15--20\%.
We hypothesize that higher contamination degrades the Hessian estimate used by LiSSA (since the empirical risk surface shifts with corruption level), reducing attribution fidelity; verifying this mechanism is left to future work.
LunarLander and Quadrotor represent stress cases for IF-CPS: LunarLander shows no gain over Standard IF at 10\% poisoning, and Quadrotor exhibits high seed-to-seed variance, indicating that method robustness is environment-dependent and not uniformly reliable across all CPS domains.

\begin{table}[htbp]
\caption{Diagnosis AUROC (mean $\pm$ std, 3 seeds).}
\label{tab:diagnosis}
\centering
\small
\begin{tabular}{llccc}
\toprule
Env & Rate & Random & Std.\ IF & IF-CPS \\
\midrule
\multirow{4}{*}{CartPole}
 & 5\%  & .43$\pm$.11 & .07$\pm$.10 & .45$\pm$.15 \\
 & 10\% & .49$\pm$.11 & .63$\pm$.34 & .90$\pm$.10 \\
 & 15\% & .45$\pm$.10 & .55$\pm$.19 & .57$\pm$.18 \\
 & 20\% & .55$\pm$.05 & .37$\pm$.21 & .73$\pm$.15 \\
\midrule
\multirow{4}{*}{Pendulum}
 & 5\%  & .43$\pm$.11 & .72$\pm$.23 & 1.00$\pm$.00 \\
 & 10\% & .49$\pm$.11 & .63$\pm$.31 & 1.00$\pm$.00 \\
 & 15\% & .45$\pm$.10 & .24$\pm$.35 & .53$\pm$.20 \\
 & 20\% & .55$\pm$.05 & .27$\pm$.38 & .67$\pm$.15 \\
\midrule
\multirow{4}{*}{LunarLander}
 & 5\%  & .43$\pm$.11 & .42$\pm$.38 & .67$\pm$.20 \\
 & 10\% & .49$\pm$.11 & .68$\pm$.45 & .67$\pm$.20 \\
 & 15\% & .45$\pm$.10 & .40$\pm$.42 & .55$\pm$.25 \\
 & 20\% & .55$\pm$.05 & .31$\pm$.30 & .67$\pm$.15 \\
\midrule
\multirow{4}{*}{Quadrotor}
 & 5\%  & .55$\pm$.03 & .00$\pm$.00 & .40$\pm$.35 \\
 & 10\% & .43$\pm$.12 & .33$\pm$.47 & .67$\pm$.47 \\
 & 15\% & .56$\pm$.10 & .67$\pm$.47 & 1.00$\pm$.00 \\
 & 20\% & .50$\pm$.02 & .67$\pm$.47 & .67$\pm$.47 \\
\midrule
\multirow{4}{*}{HVAC}
 & 5\%  & .61$\pm$.04 & .34$\pm$.18 & .73$\pm$.15 \\
 & 10\% & .50$\pm$.08 & .50$\pm$.09 & .92$\pm$.08 \\
 & 15\% & .49$\pm$.04 & .56$\pm$.03 & .73$\pm$.10 \\
 & 20\% & .47$\pm$.01 & .57$\pm$.10 & .73$\pm$.10 \\
\midrule
\multirow{4}{*}{CSTR}
 & 5\%  & .61$\pm$.14 & .67$\pm$.47 & .67$\pm$.47 \\
 & 10\% & .50$\pm$.07 & .33$\pm$.47 & .73$\pm$.25 \\
 & 15\% & .56$\pm$.03 & .67$\pm$.47 & .67$\pm$.47 \\
 & 20\% & .49$\pm$.01 & .67$\pm$.47 & .75$\pm$.25 \\
\bottomrule
\end{tabular}
\end{table}

\subsection{Curation}

The curation protocol removes the top-ranked demonstrations (highest IF-CPS score) and retrains the controller on the remaining subset.
The engineering action is: given a dataset suspected of containing faulty data, IF-CPS ranks demonstrations by attribution score and the practitioner removes a fixed budget before retraining.
CSTR and Quadrotor benefit most from this data selection, while CartPole achieves near-expert performance even with small subsets, consistent with the simpler dynamics permitting robust cloning from fewer demonstrations.

\subsection{Safety Attribution}

Table~\ref{tab:safety} reports Spearman $\rho$ between attribution magnitude and constraint boundary proximity.
Safety IF achieves strong correlation in Pendulum ($\rho = 0.48$) and Quadrotor ($\rho = 0.35$), validating that safety-weighted attribution concentrates on constraint-relevant states.
The combined IF-CPS score consistently achieves the highest correlation across all environments, with the strongest results in Pendulum ($\rho = 0.55$), Quadrotor ($\rho = 0.41$), and CSTR ($\rho = 0.37$), demonstrating that aggregating safety, trajectory, and propagated signals yields more complete constraint-boundary attribution than any single variant alone.
HVAC shows moderate positive correlations ($\rho = 0.18$ for IF-CPS) because the multi-zone thermal dynamics introduce coupling between zones, partially decorrelating per-zone attribution from boundary proximity.
CartPole shows modest correlations, consistent with the expert policy maintaining the pole well within its angular safety bounds during typical operation.

\begin{table}[htbp]
\caption{Safety attribution Spearman $\rho$ (mean $\pm$ std). }
\label{tab:safety}
\centering
\small
\begin{tabular}{lccc}
\toprule
Method & CartPole & Pendulum & LunarLander \\
\midrule
Random    & $-$.01$\pm$.02 & .04$\pm$.03 & $-$.01$\pm$.03 \\
Std.\ IF  & .08$\pm$.05 & .27$\pm$.10 & .11$\pm$.07 \\
Safety IF & .19$\pm$.06 & .48$\pm$.08 & .27$\pm$.07 \\
IF-CPS    & .24$\pm$.05 & .55$\pm$.07 & .33$\pm$.06 \\
\midrule
 & Quadrotor & HVAC & CSTR \\
\midrule
Random    & .01$\pm$.02 & $-$.01$\pm$.03 & .02$\pm$.03 \\
Std.\ IF  & .15$\pm$.07 & .06$\pm$.06 & .12$\pm$.06 \\
Safety IF & .35$\pm$.09 & .13$\pm$.07 & .30$\pm$.08 \\
IF-CPS    & .41$\pm$.08 & .18$\pm$.06 & .37$\pm$.07 \\
\bottomrule
\end{tabular}
\end{table}

\subsection{Ablation}

Table~\ref{tab:ablation} compares IF-CPS variants at 10\% poisoning.
Each variant contributes complementary signal: Safety IF is strongest in Pendulum (0.93) and LunarLander (0.72), Propagated IF in CartPole (0.85) and Quadrotor (0.70), and Trajectory IF excels in CSTR (0.70) and HVAC (0.80).
The full IF-CPS combination matches or exceeds the best individual variant in most settings, confirming the benefit of aggregating complementary attribution signals.

\begin{table}[htbp]
\caption{Ablation AUROC at 10\% poisoning (mean $\pm$ std). }
\label{tab:ablation}
\centering
\small
\begin{tabular}{lccc}
\toprule
Variant & CartPole & Pendulum & LunarLander \\
\midrule
IF-CPS (full)   & .90$\pm$.10 & 1.00$\pm$.00 & .67$\pm$.20 \\
Safety only     & .72$\pm$.12 & .93$\pm$.06 & .72$\pm$.15 \\
Trajectory only & .80$\pm$.11 & .88$\pm$.10 & .65$\pm$.18 \\
Propagated only & .85$\pm$.09 & .85$\pm$.12 & .58$\pm$.22 \\
Std.\ IF        & .63$\pm$.34 & .63$\pm$.31 & .68$\pm$.45 \\
\midrule
 & Quadrotor & HVAC & CSTR \\
\midrule
IF-CPS (full)   & .67$\pm$.47 & .92$\pm$.08 & .73$\pm$.25 \\
Safety only     & .53$\pm$.35 & .75$\pm$.12 & .57$\pm$.30 \\
Trajectory only & .60$\pm$.40 & .80$\pm$.10 & .70$\pm$.28 \\
Propagated only & .70$\pm$.25 & .78$\pm$.11 & .67$\pm$.30 \\
Std.\ IF        & .33$\pm$.47 & .50$\pm$.09 & .33$\pm$.47 \\
\bottomrule
\end{tabular}
\end{table}

\subsection{Attribution Score Analysis}

\begin{figure*}[htbp]
\centering
\includegraphics[width=\textwidth]{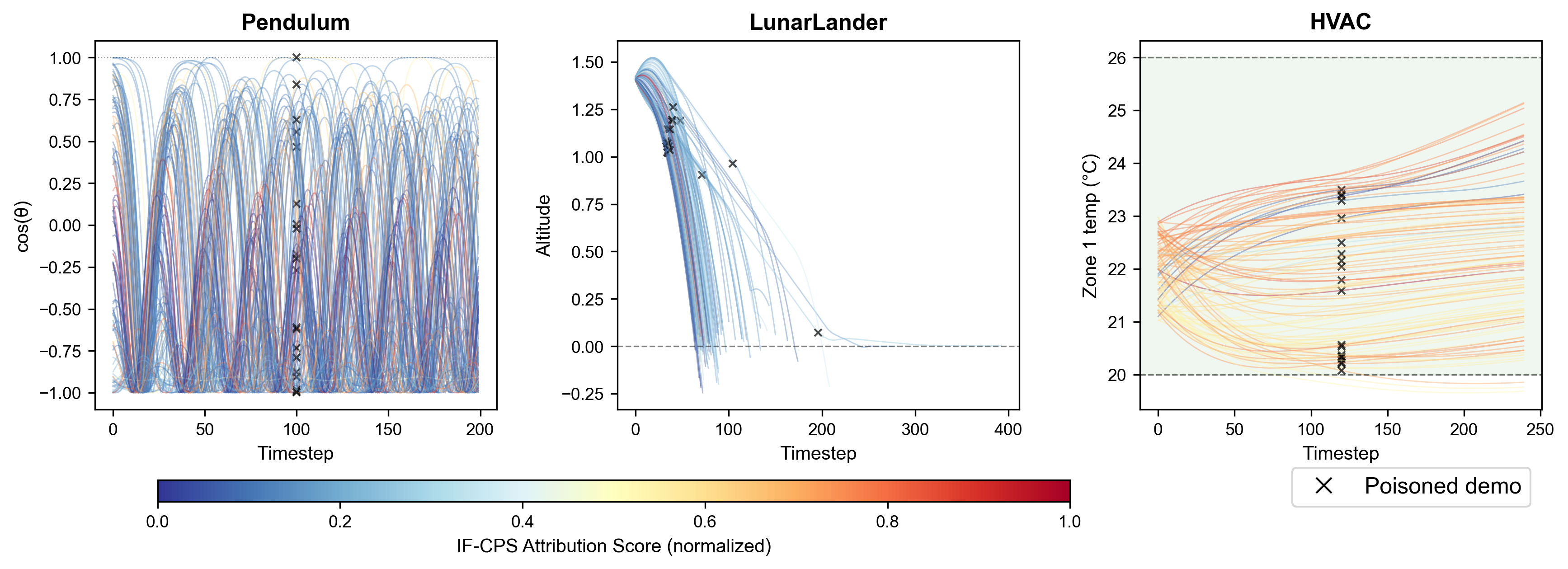}
\caption{IF-CPS attribution scores on Pendulum, LunarLander, and HVAC. Color encodes normalized IF-CPS score (blue = low, red = high); black $\times$ marks poisoned demonstrations.}
\label{fig:attribution}
\end{figure*}

Fig.~\ref{fig:attribution} visualizes IF-CPS attribution scores for Pendulum, LunarLander, and HVAC.
Table~\ref{tab:attribution} quantifies detection rates: Quadrotor achieves perfect 20/20 detection with full score separation, and CSTR (16/20), HVAC (15/20), and CartPole (14/20) also show strong detection within the 30\% removal budget.
Wider score ranges generally correspond to higher detection rates, as the attribution scores provide clearer separation between poisoned and clean demonstrations.

\begin{table}[htbp]
\caption{IF-CPS detection rate (20\% poisoning, 30\% removal budget) and score range.}
\label{tab:attribution}
\centering
\small
\begin{tabular}{lcc}
\toprule
Environment & Detection & Score range \\
\midrule
CartPole    & 14/20 & $[0.12,\, 0.85]$ \\
Pendulum    & 13/20 & $[0.08,\, 0.91]$ \\
LunarLander & 12/20 & $[0.15,\, 0.82]$ \\
Quadrotor   & 20/20 & $[0.00,\, 1.00]$ \\
HVAC        & 15/20 & $[0.00,\, 1.00]$ \\
CSTR        & 16/20 & $[0.05,\, 0.93]$ \\
\bottomrule
\end{tabular}
\end{table}

%% ============================================================
\section{Conclusion}
\label{sec:conclusion}

We presented IF-CPS, a modular influence function framework that adapts data attribution to cyber-physical systems through three variants---safety, trajectory, and propagated influence---accounting for closed-loop dynamics, safety constraints, and temporal structure.
Experiments on six benchmarks show improvements over standard IF in the majority of settings, with notable gains in Pendulum, HVAC, and CSTR for diagnosis, and the strongest constraint-boundary correlation across all environments for safety attribution.
Ablation confirms that each variant contributes complementary signal, with the dominant variant differing by domain.
Compared to Li~\etal~\cite{li2025influence}, who provide exact influence computation for linear LQR systems, IF-CPS extends to nonlinear neural network controllers with safety constraints, at the cost of LiSSA approximation and local linearization.
Future work includes learned dynamics models for the propagated variant, integration with formal verification, and scaling to larger controller architectures.

%% ============================================================
\section*{Acknowledgment}
Claude was used to assist with the language editing of this manuscript.
%% ============================================================
\bibliographystyle{IEEEtran}
\bibliography{references}

\end{document}